\documentclass[]{article}
\usepackage{epsfig}
\usepackage{cite}
\usepackage{indentfirst}
\usepackage{newlfont}
\usepackage{amsthm}
\usepackage{latexsym}
\usepackage{amsmath, amssymb, amsfonts, mathrsfs, dsfont, mathtools, empheq}
\usepackage{amsthm}
\usepackage{graphicx}
\usepackage{comment}
\usepackage{color}

\newtheorem{remark}{Remark}
\newtheorem{proposition}{Proposition}
\newtheorem{theorem}{Theorem}

\newtheorem{lemma}{Lemma}

\def\be{\begin{equation}}
\def\ee{\end{equation}}

\newcommand{\R}{\mathbb R}
\newcommand{\E}{\mathbb E}
\renewcommand{\P}{\mathbb P}
\newcommand{\1}{\mathds 1}

\newcommand{\dx}{\mathrm{d}x}
\newcommand{\dy}{\mathrm{d}y}
\newcommand{\dxdy}{\dx\,\dy}

\newcommand{\D}{\mathscr D}
\renewcommand{\O}{O}

\newcommand{\Zz}{Z^{(0)}}

\newcommand{\hess}{\mathrm{Hess}}

\newcommand{\x}{x_*}
\newcommand{\y}{y_*}

\newcommand{\K}{\mathcal K}
\renewcommand{\L}{\mathcal L}
\newcommand{\M}{\mathcal M}

\newcommand{\gau}{\phi} 

\title{Finite-size corrections for the attractive mean-field monomer-dimer model}
\author{Diego Alberici$^1$ \and Pierluigi Contucci$^1$ \and Rachele Luzi$^1$ \and Cecilia Vernia$^2$}
\date{\small $^1$Dipartimento di Matematica, Universit\`a di Bologna, Italy\\
$^2$Dipartimento di Scienze Fisiche Informatiche e Matematiche, Universit\`a di Modena e Reggio Emilia, Italy\\[2ex]
\today}

\begin{document}
\maketitle

\abstract{The finite volume correction for a mean-field monomer-dimer system with an attractive interaction are computed for the pressure density, the monomer density and the susceptibility. The results are obtained by introducing a two-dimensional integral representation for the partition function decoupling both the hard-core interaction and the attractive one. The next-to-leading terms for each of the mentioned quantities is explicitly derived as well as the value of their sign that is related to their monotonic convergence in the thermodynamic limit.}

\section{Introduction}
Monomer-dimer systems are known to have no phase transitions when the only interaction is the hard-core one. This fact 
has been rigorously proved by Heilmann and Lieb \cite{HL,HLprl}.
If instead an attractive interaction is present, favouring configurations where similar particles sit in neighbouring sites, a phase transition may be expected and has been studied for some finite-dimensional cases \cite{HLliq,A,JL} and in the mean-field setting \cite{ACM,ACMepl}, later developed and applied in \cite{ACFM,ACFMepl,Chen,CLV,SZZ}.
Monomer-dimer models are also related  to  the  matching  problem  in  computer  science, where  the  statistical mechanics  approach  has  conveyed  important  results \cite{ZM,BLS,AC}.

In the present paper we continue the investigation of the mean-field case by controlling the finite-size corrections of the main thermodynamic quantities describing the model, namely the pressure density, the monomer density and the susceptibility. This is relevant in Statistical Mechanics and its applications \cite{BCSV, ABCPV, Adriano} because the size and the sign of the corrections carry some important information on the phase transition within the phase space.

More precisely, for instance for the pressure, we are interested in proving the existence of the limit and the relative properties for the next-to-leading term:
\be
\Lambda_N \, = \, \log Z_N - N p_* \, ,
\ee
where $p_*$ is the pressure density in the thermodynamic limit.
Among the informations that $\Lambda_N$ carries, its sign is related to the type of monotonic behaviour for large $N$, namely whether the pressure reaches its limit from above or below, and is therefore important to understand whether the finite volume approximation is by excess or defect (see \cite{CLV} for its relevance in the inverse problem). 
We notice that while for the systems in finite-dimensional lattices the next to leading terms identify surface contributions and further sub-leading powers of the linear size (see for instance for the ferromagnetic Ising model \cite{FC} and spin-glass Edward-Anderson model \cite{CG}), in the mean-field case the first correction is of order one. In finite-dimensional lattices moreover the sign of the next to leading terms are related to local correlation inequalities (see \cite{Gr,Gr2,KS} for the ferromagnets and \cite{CL} for the spin-glass Edwards-Anderson), while sometimes in the mean field case there are global positivity properties that leads to positivity \cite{CDGG}.
In our case the presence of two interactions of different nature, the repulsive hard-core that forbids the overlap of two particles in the same site and the attraction that favours the closeness of similar articles, makes the identification of the next to leading term particularly challenging. To this purpose we introduce a new technical tool, a two-dimensional transform able to decouple, separately, the two interactions. Like in the case of the Laplace transform in standard ferromagnets this enable us to obtain explicit expression in terms of the solution of the model and evaluate the sign of each correction.

\section{Definitions and results}

Consider the complete graph of size $N$. A monomer-dimer configuration $D$ on the set of vertices $V_N=\{1,\dots,N\}$ is a partition into pairs of a subset $A\subset V_N\,$: the pairs $\{i,j\}\in D$ are called \textit{dimers}, while the vertices $i \in V_N\smallsetminus A$ are called \textit{monomers}. We denote the monomer density by
\be \label{eq:mN}
m_N(D) = \frac{1}{N}\,|V_N\smallsetminus A| = \frac{N-2\,|D|}{N} \;.
\ee
Beyond the hard-core interaction (two dimers cannot overlap), we consider also a mean-field attractive interaction and we define the following Hamiltonian:
\be \label{eq:H}
H_N(D) = -N\,\left(\frac{a}{2}\,m_N(D)^2 + b\,m_N(D)\right) \;,
\ee
with parameters $a>0$ and $b\in\R$.
Denoting by $\D_N$ the configuration space, the partition function of the system is
\be \label{eq:Z}
Z_N = \sum_{D\in\D_N} N^{-|D|}\, e^{-H_N(D)} \;.
\ee
We denote by $\langle\,\cdot\,\rangle_N$ the expected value with respect to the associated Gibbs measure, namely for any observable $f:\D_N\to\R$,
\be 
\langle\,f\,\rangle_N = \frac{1}{Z_N}\,\sum_{D\in\D_N} f(D)\, N^{-|D|}\, e^{-H_N(D)}
\ee
It is worth remarking that this model coincides with that studied in \cite{ACM}, by the change of parameters
\be 
a=2J\,,\quad b=h-J \;.
\ee

\begin{theorem}[Finite-size corrections] \label{th:finitesize}
Let $a>0,\,b\in\R$ such that the system has a unique phase (see \cite{ACM} for the coexistence line).
The \textit{pressure density} of a system of size $N$ is
\be \label{eq:chi_expansion}
p_N := \frac{1}{N}\,\log Z_N \,=\, p_* + \frac{\Lambda}{N} + \O\!\left(\frac{1}{N^2}\right) \;,
\ee
the \textit{average monomer density} is
\be \label{eq:m_expansion}
\mu_N := \langle\,m_N\,\rangle_N \,=\, m_* + \frac{\Lambda'}{N} + \O\!\left(\frac{1}{N^2}\right) \;,
\ee
and the \textit{susceptibility} is
\be \label{eq:chi_expansion}
\chi_N := N \left(\langle\,m_N^2\,\rangle_N - \langle\,m_N\,\rangle_N^2\right) \,=\, \chi_* + \frac{\Lambda''}{N} + \O\!\left(\frac{1}{N^2}\right) \;.
\ee
$p_*,\,m_*,\,\chi_*,\,\Lambda,\,\Lambda',\,\Lambda''$ depend on the parameters $a,b$, but not on the size $N$. Their expressions rely on the implicit expression for the limiting monomer density $m_*=\y$ (see the self-consistent equation \eqref{eq:ce2} or, equivalently refer to \cite{ACM}). In terms of $F(x,y)$, $D$, $\L$, $\K_G,\,\M_G$ that will be defined precisely in Section \ref{sec:computation} (equations \eqref{eq:F}, \eqref{eq:D}, \eqref{eq:L}, \eqref{eq:K}, \eqref{eq:M} respectively), we have: 
\begin{align}
& p_* = F\!\left(\sqrt{1- m_*}\,,m_*\right) \\
& \Lambda = - \log\sqrt{\frac{D}{a}} \\
& \Lambda' = \K_g \\
& \chi_* = m_*\,(1-m_*) + \K^{(1)}_{(g-m_*)^2} \label{eq:chi*} \\
& \Lambda'' = \K_{g\,(1-g)} - \K^{(1)}_{(g-m_*)^2}\,\L + \M_{(g-m_*)^2} - (\K_g)^2
\end{align}
\end{theorem}

The computation of the finite size corrections relies on the following integral representation, which decouples both the attractive interaction and the hard-core interaction.

\begin{proposition}[Integral representation] \label{prop:representation}
For any $a>0,\,b\in\R$, the partition function admits the following integral representation:
\be \label{eq:integral}
Z_N = \frac{N\sqrt{a}}{2\pi}\, \iint_{\R^2} \Phi(x,y)^N \,\dxdy \;,
\ee
where
\be \label{eq:phi}
\Phi(x,y) \,=\, \left(x+e^{ay+b}\right)\,\exp\!\left(-\frac{x^2}{2}-a\,\frac{y^2}{2}\right) \;.
\ee
\end{proposition}

The previous integral representation is based on two properties of Gaussian measures, that we recall in the next lemmas.

\begin{lemma}[Hubbart-Stratonovich transform, \textit{or} the Gaussian moments generating function]
For any $\sigma>0$ and $t\in\R$,
\be \label{eq:HS}
\exp\left(\frac{t^2\sigma^2}{2}\right) \,=\, \frac{1}{\sqrt{2\pi\sigma^2}}\, \int_\R \exp\left(t\,y-\frac{y^2}{2\sigma^2}\right) \,\dy
\ee
\end{lemma}

\begin{lemma}[Wick-Isserlis rule for the Gaussian moments]
For any $\sigma>0$ and any finite set $A$,
\be \label{eq:Wick}
\sum_{D\textrm{ partition of $A$ into pairs}} \sigma^{2|D|} \,=\, \frac{1}{\sqrt{2\pi\sigma^2}}\, \int_\R x^{|A|} \exp\left(-\frac{x^2}{2\sigma^2}\right) \,\dx
\ee
\end{lemma}

\proof[Proof of Proposition \ref{prop:representation}]
First we use the Hubbart-Stratonovich transform to decouple the attractive interaction. Choosing $t=Na\,m_N(D)$ and $\sigma^2 = \frac{1}{Na}$ in \eqref{eq:HS}, the partition function \eqref{eq:Z} rewrites as
\be \label{eq:attractive_decouple} \begin{split}
Z_N(a,b) & = \sum_{D\in\D_N} N^{-|D|} \exp N\!\left(\frac{a}{2}\,m_N(D)^2 + b\,m_N(D)\right) = \\
& = \sum_{D\in\D_N} N^{-|D|} \sqrt{\frac{Na}{2\pi}} \int_\R \exp N\!\left((ay+b)\,m_N(D) -\frac{a y^2}{2}\right) \dy = \\
& = \sqrt{\frac{Na}{2\pi}} \int_\R \Zz_N(ay+b)\; \exp\!\left(- N\frac{a y^2}{2}\right) \,\dy
\end{split} \ee
where $\Zz_N(b')$ denotes the partition function $Z_N(a=0,\,b=b')$ of the model without attractive interaction.
Now we use the Wick-Isserlis rule to decouple the hard-core interaction, as was shown in \cite{ACMrand,Vlad}. Choosing $\sigma^2= N^{-1}$ and $A\subseteq V_N$ in \eqref{eq:Wick}, the non-attractive partition function rewrites as
\be \label{eq:hardcore_decouple} \begin{split}
\Zz_N(b') & = \sum_{D\in\D_N} N^{-|D|} \exp\!\left(b'Nm_N(D)\right) = \\
& = \sum_{A\subseteq V_N} e^{b'(N-|A|)}\, \sqrt{\frac{N}{2\pi}}\, \int_\R x^{|A|} \exp\left(-N\frac{x^2}{2}\right) \dx = \\
& = \sqrt{\frac{N}{2\pi}}\, \int_\R (x+e^{b'})^N \exp\left(-N\frac{x^2}{2}\right) \dx \;.
\end{split}\ee 
Substituting \eqref{eq:hardcore_decouple} with $b'=ay+b$ into \eqref{eq:attractive_decouple}, we finally obtain \eqref{eq:integral}.
\endproof

\section{Monotonicity regions}
\begin{figure}
\centering
\includegraphics[width=\textwidth]{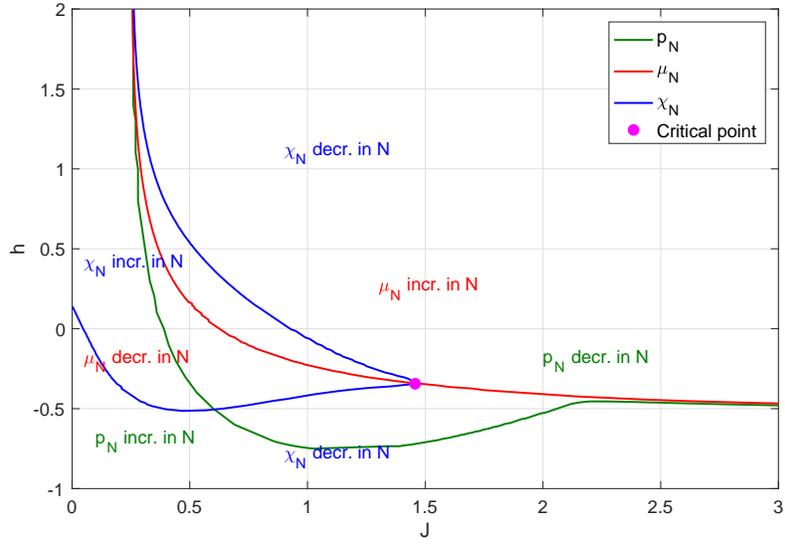}
\caption{\label{fig1} \small Phase space $(J,h)$. The green curve represents the values $(J,h)$ for which the 
pressure density $p_N$ changes monotonicity with respect to $N$ (i.e. $\Lambda$ changes sign).
The red curve represents the values $(J,h)$ for which the 
average monomer density $\mu_N$ changes monotonicity with respect to $N$ (i.e. $\Lambda'$ changes sign).
The blue curve represents the values $(J,h)$ for which the 
susceptibility $\chi_N$ changes monotonicity with respect to $N$ (i.e. $\Lambda''$ changes sign).
The purple dot is the critical point of the system.}
\end{figure}

The finite-size corrections for the \textit{pressure density} $p_N$, the \textit{average monomer density} $\mu_N$ and the \textit{susceptibility} $\chi_N$ can be used to determine the monotonicity of the three sequences with respect to the size of the system $N$.
To be precise, the signs of the corrections $\Lambda,\,\Lambda',\,\Lambda''$ in Theorem \ref{th:finitesize} determine whether $p_N,\,\mu_N,\,\chi_N$ reach their respective limits $p_*,\,m_*,\,\chi_*$ from above or from below.
Figure \ref{fig1} shows the phase space regions where $\Lambda$ (green curve), $\Lambda'$ (red curve) and $\Lambda''$ (blue curve) change sign.

\begin{figure}
\centering
\includegraphics[width=\textwidth]{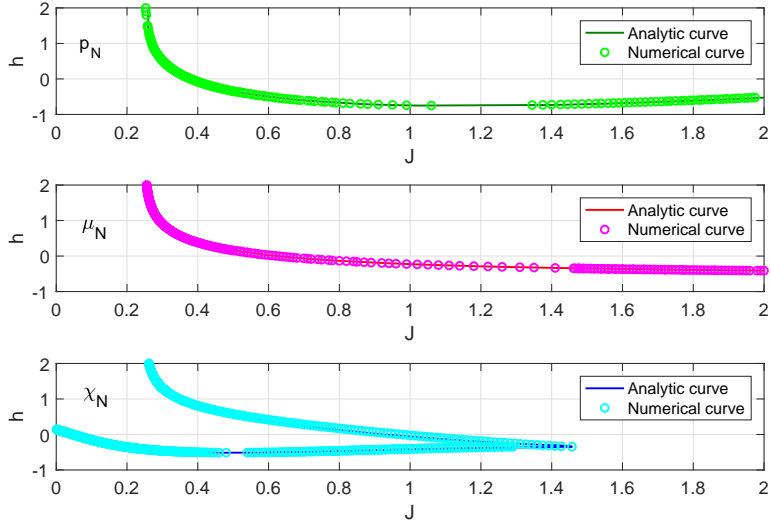}
\caption{\label{fig2} \small Phase space $(J,h)$: comparison between analytical and numerical monotonicity (with respect to $N$) curves for the pressure density $p_N$ (upper panel), for the average monomer density $\mu_N$ (middle panel) and for the susceptibility $\chi_N$ (lower panel). The analytical curves (continuous lines) are the same as in Fig. \ref{fig1}.}
\end{figure}

Due to the mean field nature of the model, the Gibbs measure and the expected value with respect to such measure at finite volume size $N$ can be computed by evaluating the combinatorial weights of the possible dimer density values, that is the number of the possible configurations that share the same value $|D|$ of dimers on the complete graph with $N$ vertices \cite{CLV}.
This enable us to compute numerically the phase space curves in which $p_N,\,\mu_N$ and $\chi_N$ invert the monotonicity with respect to $N$. The comparison of these numerical curves with the ones obtained analitically from $\Lambda$, $\Lambda'$ and $\Lambda''$ (Fig. \ref{fig1}) is shown in Fig. \ref{fig2}.
The perfect overlap between the curves is evident.

\begin{remark}
It is possible to modify the Hamiltonian of the system in such a way that the Gibbs measure does not change, but the new pressure density $\tilde p_N$ reaches its limit $p_*$ from above as $N\to\infty$ in the whole phase space $(a,b)$. Namely
\be 
\lim_{N\to\infty} \tilde p_N = \inf_{N} \tilde p_N \quad\forall a\geq0,\,b\in\R \;.
\ee
The finite size correction $\Lambda=-\log\sqrt{D/a}\,$ has range $(-\log\sqrt2,+\infty)$, indeed it is easy to compute explicitly the determinant
\be
\frac{D}{a} \,=\, 2-\y-2\,a\,\y(1-\y)
\ee
which is non-negative by definition, takes value $0$ at the critical point of the system and goes to $2$ as $b\to-\infty$ and $a$ is fixed.
Therefore it suffices to set $\tilde H_N(D) = H_N(D) - \log\sqrt 2$ for all configurations $D\in\D_N$ in order to obtain
\be 
\tilde p_N \,=\, p_N + \frac{\log\sqrt{2}}{N} \,=\, p_* + \frac{\tilde \Lambda}{N} + \O\left(\frac{1}{N^2}\right)
\ee
where $\tilde\Lambda = \Lambda + \log\sqrt{2}\,>0\,$ for all $a\geq0,\,b\in\R\,$.

On the contrary it is not possible to obtain a modified pressure density reaching $p_*$ from below in the whole phase space, since the upper bound of the finite size correction $\Lambda$ is $+\infty$.
\end{remark}

\section{Computation of the finite-size corrections} \label{sec:computation}
The integral representation \eqref{eq:integral} allows to compute the finite-size corrections by estimations of suitable Laplace two-dimensional integrals \cite{DeBru}. Laplace estimates are needed up to order $N^{-1}$ in the case of the magnetisation and up to order $N^{-2}$ in the case of the susceptibility. 

We denote by $(\x,\y)$ the global maximum of $\Phi$ on $\R^2$. First of all, we observe that $\x,\y>0$ since
\be 
\Phi(s_1 x,s_2 y) \leq \Phi(x,y)\quad \forall x,y>0\quad \forall s_1,s_2=\pm1
\ee 
and the inequality is strict if $s_1,s_2$ are not both $1$.
It is convenient to set for $x,y>0$
\be \label{eq:F}
F(x,y) := \log \Phi(x,y) = -\frac{x^2}{2} - a\,\frac{y^2}{2} + \log\left(x+e^{ay+b}\right) \;.
\ee
The condition $\nabla F(\x,\y)=0$ says that $(\x,\y)$ is a solution of the following fixed point system:
\be \label{eq:ce1}
\begin{cases}x= \dfrac{1}{x+e^{ay+b}} \\[10pt] y= \dfrac{e^{ay+b}}{x+e^{ay+b}} \end{cases}
\ee
which rewrites as:
\be \label{eq:ce2}
\begin{cases} x=\sqrt{1-y} \\[4pt] e^{ay+b} = \dfrac{y}{\sqrt{1-y}} \end{cases} \;.
\ee
The second equation in \eqref{eq:ce2} is self-consistent and an elementary analysis shows that it has a unique solution $\y$ for $a\leq a_c := (3+2\sqrt2)/2$. For $a>a_c$ there are at most $3$ solutions
, one in each of the following intervals $(0,y_-)$, $(y_-,y_+)$, $(y_+,1)$, where $y_\pm := \left(2 a + 1 \pm \sqrt{4 a^2 - 12 a + 1}\right) \,/\, (4a)\,$;
$\y$ is the solution maximizing $F(\x,\y)$ with $\x=\sqrt{1-\y}$.
$\y$ will be two-values only for parameters $a,b$ on the coexistence line \cite{ACM}, a case that we exclude from the present paper for the sake of simplicity.

Now set also
\be 
g(x,y) := \frac{\partial F}{\partial b}(x,y) = \frac{e^{ay+b}}{x+e^{ay+b}} \;.
\ee
The \textit{average monomer density}, using the integral representation \eqref{eq:integral} for the partition function, rewrites as
\be \label{eq:m_integral}
\mu_N = \frac{1}{N}\frac{\partial}{\partial b}\log Z_N = g(\x,\y) \,+\, \frac{\displaystyle\iint \Phi(x,y)^N \left(g(x,y)-g(\x,\y)\right)\,\dxdy}{\displaystyle\iint \Phi(x,y)^N\,\dxdy} 
\ee
The \textit{susceptibility} rewrites as
\be \label{eq:chi_integral}\begin{split}
\chi_N & = \frac{1}{N}\frac{\partial^2}{\partial b^2}\log Z_N = \\
& = N\,\frac{\displaystyle\iint \Phi(x,y)^N \left(g(x,y)-g(\x,\y)\right)^2\,\dxdy}{\displaystyle\iint \Phi(x,y)^N\,\dxdy}\, + \\ & \phantom{=} - N \left(\frac{\displaystyle\iint \Phi(x,y)^N \left(g(x,y)-g(\x,\y)\right)\,\dxdy}{\displaystyle\iint \Phi(x,y)^N\,\dxdy}\right)^{\!2} + \\ & \phantom{=} + \frac{\displaystyle\iint \Phi(x,y)^N\,g(x,y)\left(1-g(x,y)\right)\,\dxdy}{\displaystyle\iint \Phi(x,y)^N\,\dxdy}
\end{split} \ee
Both in \eqref{eq:m_integral} and \eqref{eq:chi_integral}, the term $g(\x,\y)$ has been artificially introduced in order to simplify the following computations.
By the way, observe that according to \eqref{eq:ce1}, $g(\x,\y)=\y\,$.

Expressions \eqref{eq:m_integral} and \eqref{eq:chi_integral} can be typically approximated by the Laplace method. These estimates involve Gaussian moments and higher order derivatives of $F$ and $g$ at the maximum point $(\x,\y)$ of $F$.
Therefore it will be convenient to introduce the following notations:
\be 
F_{i,j} := \frac{\partial^{i+j}F}{\partial x^i\partial y^j}(\x,\y) \;,
\ee
while
\be
\gau_{i,j} := \frac{1}{2\pi}\,\iint_{\R^2} x^i y^j \exp\left(-\frac{1}{2}\,(x,y)C(x,y)^T\right) \dxdy
\ee
where
\be \label{eq:C}
C:= \left(-\hess F(\x,\y)\right)^{-1} = \frac{1}{D}\,\begin{pmatrix} -F_{0,2} & F_{1,1} \\ F_{1,1} & -F_{2,0} \end{pmatrix}  \;,
\ee
\be \label{eq:D}
D := \det(-\hess F(\x,\y)) = F_{0,2}F_{2,0} - F_{1,1}^2 \;.
\ee

\begin{proposition}[Laplace estimates] \label{prop:laplace}
Consider the integral
\be 
I_N(G) := \iint_{\R^2} \Phi(x,y)^N \,G(x,y)\, \dxdy \;,
\ee
where $G$ is any real function, analytic in a neighbourhood of $(\x,\y)$. Set
\be 
I_N'(G) :=  I_N(G) \Big/  \left(\frac{2\pi}{N\sqrt{D}}\, e^{NF(\x,\y)}\right) \;.
\ee
Provided that the global maximum point $(\x,\y)$ is unique and the Hessian matrix $-C^{-1}$ is negative definite, the following estimates hold true:
\begin{itemize}
\item[a)] \be
I_N'(G) \,=\, G(\x,\y) + \O\left(N^{-1}\right)
\ee
\item[b)] If $G(x,y)\equiv 1$ for all $(x,y)\in\R^2$,
\be 
I_N'(G) \,=\, 1 + \L\,N^{-1} + \O\left(N^{-2}\right) \,
\ee
where
\be \label{eq:L} \begin{aligned}
& \L = \L^{(1)} + \L^{(2)} \;,\\
&\L^{(1)} = \sum_{i+j=4} \frac{F_{i,j}}{i!j!}\,\frac{\gau_{i,j}}{D} \\
&\L^{(2)} = \sum_{\substack{i_1+j_1=3,\\i_2+j_2=3}} \frac{F_{i_1,j_1}}{i_1!j_1!}\,\frac{F_{i_2,j_2}}{i_2!j_2!}\,\frac{\gau_{i_1+i_2,j_1+j_2}}{D^2} \\
\end{aligned} \ee
\item[c)] If $G(\x,\y)=0$,
\be 
I_N'(G) \,=\, \K_G \, N^{-1} + \O\left(N^{-2}\right) \;,
\ee
where
\be \label{eq:K} \begin{aligned}
& \K_G = \K^{(1)}_G+ \K^{(2)}_G \;,\\
&\K^{(1)}_G = \sum_{i+j=2} \frac{G_{i,j}}{i!j!}\,\frac{\gau_{i,j}}{D} \\
&\K^{(2)}_G = \sum_{\substack{i_1+j_1=3,\\i_2+j_2=1}} \frac{F_{i_1,j_1}}{i_1!j_1!}\,\frac{G_{i_2,j_2}}{i_2!j_2!}\,\frac{\gau_{i_1+i_2,j_1+j_2}}{D^2} \\
\end{aligned} \ee
\item[d)] If $G(\x,\y)=0$ and $\nabla G(\x,\y)=0$,
\be 
I_N'(G) \,=\, \K^{(1)}_G\, N^{-1} + \M_G\, N^{-2} + \O\left(N^{-3}\right) \;,
\ee
where
\be \label{eq:M} \begin{aligned}
&\M_G = \M^{(1)}_G + \M^{(2)}_G + \M^{(3)}_G + \M^{(4)}_G \;,\\
&\M^{(1)}_G = \sum_{i+j=4} \frac{G_{i,j}}{i!j!}\,\frac{\gau_{i,j}}{D^2} \\
&\M^{(2)}_G = \sum_{\substack{i_1+j_1 = 3,\\i_2+j_2 = 3}} \frac{F_{i_1,j_1}}{i_1!j_1!}\,\frac{G_{i_2,j_2}}{i_2!j_2!}\,\frac{\gau_{i_1+i_2,j_1+j_2}}{D^3} \\
&\M^{(3)}_G = \sum_{\substack{i_1+j_1 = 4,\\i_2+j_2 = 2}} \frac{F_{i_1,j_1}}{i_1!j_1!}\,\frac{G_{i_2,j_2}}{i_2!j_2!}\,\frac{\gau_{i_1+i_2,j_1+j_2}}{D^3} \\
&\M^{(4)}_G = \frac{1}{2}\,\sum_{\substack{i_1+j_1 = 3,\\i_2+j_2 = 3,\\i_3+j_3=2}} \frac{F_{i_1,j_1}}{i_1!j_1!}\, \frac{F_{i_2,j_2}}{i_2!j_2!}\, \frac{G_{i_3,j_3}}{i_3!j_3!}\,\frac{\gau_{i_1+i_2+i_3,j_1+j_2+j_3}}{D^4} \;.\\
\end{aligned}\ee
\end{itemize}
\end{proposition}

\proof[Sketch of the Proof]
Since $\Phi(x,y)$ takes its maximum on $(0,1)^2$, any contribution to the integral $I_N$ coming from $(x,y)\in\R^2\smallsetminus (0,1)^2$ is exponentially small compared to the contribution given by $(x,y)\in(0,1)^2\,$. We write $I_N'\approx J_N'$ if there exists $\delta>0$ such that $I_N' = J_N' + \O(e^{-\delta N})$ for every $N$.
Observe that
\be
I_N' \,\approx\, \frac{N\sqrt{D}}{2\pi}\,\int_0^1\!\!\int_0^1\exp N\!\left(F(x,y)-F(\x,\y)\right)\, G(x,y)\,\dxdy \;.
\ee
Then make the change of variable $(x,y)\mapsto(\x,\y)+\frac{1}{\sqrt N}(x,y)$ and expand $F$ around $(\x,\y)$. Since $\nabla F(\x,\y)=0$, one obtains that
\be \label{eq:laplaceproof1}
I_N' \,\approx\, \frac{\sqrt{D}}{2\pi}\,\iint_{A_N} \exp\left(-\frac{1}{2}\,(x,y)\,C^{-1}(x,y)^T\right)\, e^{f_N(x,y)}\, G_N(x,y)\,\dxdy \;,
\ee
where $A_N := \big(-\x\sqrt N,\,(1-\x)\sqrt N\,\big)\times\big(-\y\sqrt N,\,(1-\y)\sqrt N\,\big)$ and
\be
f_N(x,y) := \sum_{i+j\geq 3} \frac{F_{i,j}}{i!j!}\,x^i y^j\,N^{-\frac{i+j}{2}+1} \;,
\ee
\be 
G_N(x,y) := \sum_{i+j\geq 0} \frac{G_{i,j}}{i!j!}\,x^i y^j\,N^{-\frac{i+j}{2}} \;.
\ee 
Let $(X,Y)$ be a centred Gaussian vector with covariance matrix $C$: \eqref{eq:laplaceproof1} rewrites as
\be 
I_N' \,\approx\, \E\left[ e^{f_N(X,Y)}\,G_N(X,Y)\; \1\left((X,Y)\in A_N\right) \right] \;.
\ee
We remark that there exist $K,\delta>0$ such that $\P\big((X,Y)\notin A_N\big)\leq K\,e^{-\delta N}$ for all $N$. Therefore the orders $1,\,N^{-1},\,N^{-2},\dots$ of $I_N'$ are obtained by multiplying the suitable terms in the Taylor expansions of $e^{f_N} = 1+f_N+\frac{1}{2}f_N^2+\dots$ and $G_N$ and computing the corresponding Gaussian moments of $(X,Y)$.
It is worth noticing that the fractional orders $N^{-\frac{1}{2}},\,N^{-\frac{3}{2}},\,\dots$ are zero because the odd Gaussian moments are zero.
\endproof

The integral representations of the pressure density \eqref{eq:integral}, the average monomer density \eqref{eq:m_integral} and the susceptibility \eqref{eq:chi_integral} can be estimated according to Proposition \ref{prop:laplace}, yielding a proof of Theorem \ref{th:finitesize}.
In the Appendix we compute explicitly the Gaussian moments and the derivatives that appear in the expressions of the finite-size corrections.

\begin{remark}
An elementary computation shows that the susceptibility limit $\chi_*$ found in the present paper \eqref{eq:chi*} coincides with that obtained in \cite{CLV} (equation 5) by direct differentiation of the consistency equation, namely
\be 
\chi_* \,=\, \frac{2\,m_* (1-m_*)}{2-m_*-2a\,m_* (1-m_*)} \;.
\ee
\end{remark}

\section*{Appendix}
The even moments of $(X,Y)$, centered Gaussian vector of covariance matrix \eqref{eq:C}, are computed up to order $8$ using the Wick's rule:
\be \begin{aligned}
&\gau_{2,0} = -F_{0,2} \,,\quad
\gau_{1,1} = F_{1,1} \,,\quad
\gau_{0,2} = -F_{2,0} \,,\\
&\gau_{4,0} = 3\,F_{0,2}^2 \,,\quad
\gau_{3,1} = -3\,F_{0,2}\,F_{1,1} \,,\\
&\gau_{2,2} = F_{0,2}\,F_{2,0} + 2\,F_{1,1}^2 \,,\\
&\gau_{1,3} = -3\,F_{1,1}\,F_{2,0} \,,\quad
\gau_{0,4} = 3\,F_{2,0}^2 \,,\\
&\gau_{6,0} = -15\,F_{0,2}^3 \,,\quad
\gau_{5,1} = 15\,F_{0,2}^2\,F_{1,1} \,,\\
&\gau_{4,2} = -3\,F_{0,2}^2\,F_{2,0} - 12\,F_{0,2}\,F_{1,1}^2 \,,\\
&\gau_{3,3} = 9\,F_{0,2}\,F_{1,1}\,F_{2,0} + 6\,F_{1,1}^3 \,,\\
&\gau_{2,4} = -3\,F_{0,2}\,F_{2,0}^2 - 12\,F_{1,1}^2\,F_{2,0} \,,\\
&\gau_{1,5} = 15\,F_{1,1}\,F_{2,0}^2 \,,\quad
\gau_{0,6} = -15\,F_{2,0}^3 \,,\\
&\gau_{8,0} = 105\,F_{0,2}^4 \,,\quad
\gau_{7,1} = -105\,F_{0,2}^3\,F_{1,1} \,,\\
&\gau_{6,2} = 90\,F_{0,2}^2\,F_{1,1}^2 + 15\, F_{0,2}^3\,F_{2,0} \,,\\
&\gau_{5,3} = -45\,F_{0,2}^2\,F_{1,1}\,F_{2,0} - 60\,F_{2,0}\,F_{1,1}^3 \,,\\
&\gau_{4,4} = 9\,F_{0,2}^2\,F_{2,0}^2 + 24\,F_{1,1}^4 + 72\,F_{0,2}\,F_{1,1}^2\,F_{2,0} \,,\\
&\gau_{3,5} = - 45\,F_{0,2}\,F_{1,1}\,F_{2,0}^2 - 60\,F_{0,2}\,F_{1,1}^3 \,,\\
&\gau_{2,6} = 90\,F_{1,1}^2\,F_{2,0}^2 + 15\,F_{0,2}\,F_{2,0}^3 \,,\\
&\gau_{1,7} = - 105\,F_{2,0}^3\,F_{1,1} \,,\quad
\gau_{0,8} = 105\,F_{2,0}^4 \,.
\end{aligned} \ee
The derivatives of $F$ at its maximum point $(\x,\y)$ up to order $4$, in terms of $\y$ only are:
\be \begin{aligned}
&F_{2,0} = -2 + \y \,,\quad
F_{1,1} = -a\, \y \sqrt{1 - \y} \,,\\
&F_{0,2} = -a + a^2\,\y (1-\y) \,,\\
&F_{3,0} = 2 (1 - \y)^{3/2} \,,\quad
F_{2,1} = 2 a\, \y (1 - \y) \,,\\
&F_{1,2} = -a^2\, \y \sqrt{1 - \y}\, (1 - 2 \y) \,,\quad
F_{0,3} = a^3\, \y (1 - \y) (1 - 2 \y) \,,\\
&F_{4,0} = -6\,(1-\y)^2 \,,\quad
F_{3,1} = -6a\,\y (1-\y)^{3/2} \,,\\
&F_{2,2} = 2a^2\, \y (1-\y) (1-3\y) \,,\\
&F_{1,3} = -a^3\, \y \sqrt{1-\y}\, (1-6\y+6\y^2) \,,\\
&F_{0,4} = a^4\, \y (1-\y) (1-6\y+6\y^2) \,.
\end{aligned}
\ee
The derivatives of $g$ at $(\x,\y)$ up to order $3$:
\be \begin{aligned}
&g_{1,0} = -\y \sqrt{1 - \y} \,,\quad
g_{0,1} = a\, \y (1 - \y) \,,\quad
g_{2,0} = 2\, \y (1 - \y) \,,\\
&g_{1,1} = -a\, \y \sqrt{1 - \y}\, (1 - 2 \y) \,,\quad
g_{0,2} = a^2\, \y (1 - \y) (1 - 2 \y) \,,\\
&g_{3,0} = -6\,\y (1-\y)^{3/2} \,,\quad
g_{2,1} = 2a\, \y (1-\y) (1-3\y) \,,\\
&g_{1,2} = -a^2\, \y \sqrt{1-\y}\, (1-6\y+6\y^2) \,,\\
&g_{0,3} = a^3\, \y (1-\y) (1-6\y+6\y^2) \,.
\end{aligned} \ee
The derivatives of $\tilde g := (g - \y)^2$ at $(\x,\y)$ up to order $4$:
\be \begin{aligned}
&\tilde g_{2,0} = 2\, g_{1,0}^2 \,,\quad
\tilde g_{1,1} = 2\,g_{1,0}\,g_{0,1} \,,\quad
\tilde g_{0,2} = 2\, g_{0,1}^2 \,,\\
&\tilde g_{3,0} = 6\,g_{1,0}\,g_{2,0} \,,\quad
\tilde g_{2,1} = 4\,g_{1,0}\,g_{1,1} + 2\,g_{0,1}\,g_{2,0} \,,\\
&\tilde g_{1,2} = 4\,g_{0,1}\,g_{1,1} + 2\,g_{1,0}\,g_{0,2} \,,\quad
\tilde g_{0,3} = 6\,g_{0,1}\,g_{0,2} \,,\\
&\tilde g_{4,0} = 6\,g_{2,0}^2 + 8\,g_{1,0}\,g_{3,0} \,,\\
&\tilde g_{3,1} = 6\,g_{2,0}\,g_{1,1} + 6\,g_{1,0}\,g_{2,1} + 2\,g_{0,1}\,g_{3,0} \,,\\
&\tilde g_{2,2} = 4\,g_{1,1}^2 + 2\,g_{2,0}\,g_{0,2} + 4\,g_{1,0}\,g_{1,2} + 4\,g_{0,1}\,g_{2,1}  \,,\\
&\tilde g_{1,3} = 6\,g_{1,1}\,g_{0,2} + 6\,g_{0,1}\,g_{1,2} + 2\,g_{1,0}\,g_{0,3} \,,\\
&\tilde g_{0,4} = 6\,g_{0,2}^2 + 8\,g_{0,1}\,g_{0,3} \,,\\
\end{aligned} \ee
The derivatives of $\hat g := g(1-g)$ at $(\x,\y)$ up to order $2$:
\be \begin{aligned}
&\hat g_{1,0} = (1-2\y)\,g_{1,0} \,,\quad
\hat g_{0,1} = (1-2\y)\,g_{0,1} \,,\\
&\hat g_{2,0} = -2\,g_{1,0}^2 + (1-2\y)\,g_{2,0} \,,\\
&\hat g_{1,1} = -2\,g_{1,0}\,g_{0,1} +(1-2\y)\,g_{1,1} \,,\\
&\hat g_{0,2} = -2\,g_{0,1}^2 + (1-2\y)\,g_{0,2} \,.
\end{aligned} \ee

$ $\\

\noindent\textbf{Acknowledgements}. The authors thank Emanuele Mingione for many useful discussions in particular those about the integral representation of the partition function. DA acknoledges financial support from Progetto Giovani 2017 GNFM Indam. PC acknowledges financial support from PRIN (grant number 2010HXAW77). CV acknowledges financial support from Fondo di Ateneo per la Ricerca 2017 (UniMoRe).

\end{document}